# The Scientific Need for a Scalar/Higgs Factory[*]

D.B. Cline

*Center for Advanced Accelerators*
*Department of Physics and Astronomy, Box 951547*
*University of California, Los Angeles, CA 90095-1547 USA*

**Abstract.** The scalar sector of the electroweak theory can be probed by a $\mu^+\mu^-$ collider S channel resonance machine. We give arguments for when such a machine may be needed and when this information could be obtained by the LHC detector. A very interesting case is the possibility that several scalar particles are in the same mass range for the supersymmetric Higgs bosons h, H, and A, which would definitely require such a machine. The Higgs factory could follow the construction of a neutrino factory.

## EARLY STUDY OF THE HIGGS FACTORY

The concept of a Higgs factory $\mu^+\mu^-$ collider was born at the first dedicated $\mu^+\mu^-$ workshop in Napa, California, December 1992;[1,2] Figure 1 shows a schematic of the scan for the Higgs presented at that meeting.[1] The next workshop also changed the role of a $\mu^+\mu^-$ collider.[3] Subsequently very nice theoretical work on this issue has been carried out by Berger, Barger, Gunion, and Han,[4] and then a paper (which included Table 1) defining the Higgs particle factory was presented at the Snowmass DPF meeting in 1996 by the author.[5] At that 1996 DPF meeting, a first pass design of a $\mu^+\mu^-$ collider was presented by the Muon Collider Consortium.[6] We will make use of all of these materials and of more recent work in this brief article, as well as proceedings from meetings we have organized[2,3,7] and other recent workshops.

When the Higgs factory was envisioned in 1992, there seemed to be little scientific support.[1]

## MOTIVATION FOR A HIGGS FACTORY

The major purpose of the Higgs factory is to find the exact Higgs mass (or masses) and then measure the important parameters, such as the width(s) and the common and rare branching fractions. This concept is based mainly on a relatively low-mass Higgs (below 300 GeV). In the low-mass region (below 150 GeV), the Higgs could well be supersymmetric (SUSY), and the width measurement will be crucial. Above 150 GeV, the Higgs could be more of a standard-model type. However, this will once again lead to the issue of what keeps the scalar system stable, which might be answered by the study of rare

---

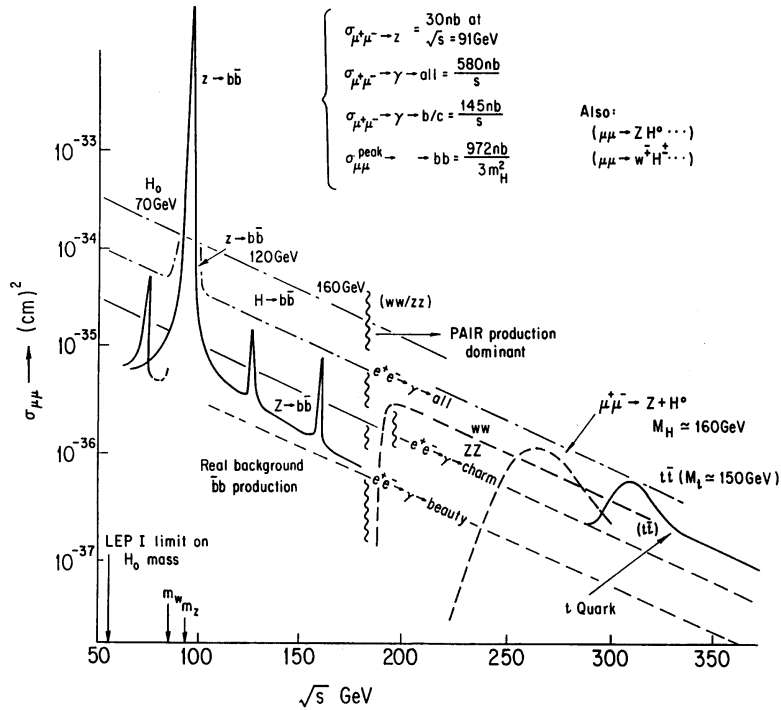

**FIGURE 1.** The first concept of a Higgs factory $\mu^+\mu^-$ collider from the Napa Workshop.[1]

**TABLE 1.** Arguments for a Higgs-Factory $\mu^+\mu^-$ Collider.[1,5]

1. The $m_\mu \backslash m_e$ ratio gives coupling 40,000 times greater to the Higgs particle. In the SUSY model, one Higgs $m_h < 120$ GeV!!
2. The low radiation of the beams makes precision energy scans possible.
3. The cost of a "custom" collider ring is a small fraction of the $\mu^\pm$ source.
4. Feasibility report to Snowmass established that $\mathcal{L} \sim 10^{33}$ cm$^{-2}$ s$^{-1}$ is feasible.

decays of the Higgs particles (in progress). In the near future, there could be evidence for the Higgs mass obtained from precise electroweak parameter measurements and later from the LHC. This will be a crucial input for the development of the Higgs factory. In addition, if Nature is supersymmetric, there will be additional SUSY-Higgs particles to study and, thus, the Higgs factory concept will include the search for and study of the SUSY Higgs ($H,A$ ...). This is an experimental issue – theory can only take us just so far!

From all we now know about elementary particle physics, the scalar or SUSY scalar sector is the key to future understanding. A complete understanding of this sector is really the goal of the Higgs factory and of nearly all elementary particle physics these days.

The Higgs factory is designed to first give the exact Higgs mass using an energy scan and then measure the general properties of the Higgs, such as the field width, largest branching fractions, *etc*. It would produce $10^4$ Higgs/yr and could investigate rare branching modes.

If there are more Higgs, the Higgs factory would be used to scan and find and study these in detail as well. A $\mu^+\mu^-$ collider provides the Higgs factory, since scalars couple like $m_l^2$ and the collider has little radiative energy spreading (see Table 2). Some important meetings are listed in Refs. 7–9, and some references to the early concept of a $\mu^+\mu^-$ collider can be found in Ref. 10.

We expect the supercollider LHC to extract the signal from background (*i.e.*, seeing either $h^0 \to \gamma\gamma$ or the very rare $h^0 \to \mu\mu\mu\mu$ in this mass range, since $h \to b\bar{b}$ is swamped by hadronic background). However, detectors for the LHC are designed to extract this signal. Figure 2 gives a picture of the various physics thresholds that may be of interest for a $\mu^+\mu^-$ collider. In this low mass region, the Higgs is also expected to be a fairly narrow resonance and, thus, the signal should stand out clearly from the background from

$$\mu^+\mu^- \to \gamma \to b\bar{b} \to Z_{tail} \to b\bar{b} \quad . \tag{1}$$

For masses above 180 GeV, the dominant Higgs decay is

$$h^0 \to W^+W^- \quad \text{or} \quad Z^0Z^0 \quad , \tag{2}$$

and the LHC should easily detect this Higgs particle. Thus the $\mu^+\mu^-$ collider is better adapted for the low mass region.

The strongest argument for the low-energy collider comes from the growing evidence that the Higgs should exist in this low-mass range from recent electroweak studies as shown on Fig. 3.[11] This evidence implies the exciting possibility that the Higgs mass is just beyond the reach of LEP II and in a range that is very difficult for the LHC to detect.[3]

In Fig. 2, we show a comparison of the Higgs factory $\mu^+\mu^-$ collider and an $e^+e^-$ collider (NLC) that could also study the Higgs.[5] Note the very great differences in cross sections, indicating that the $e^+e^-$ collider must have very high luminosity.

**TABLE 2.** Logic of Detailed Study of the Higgs Sector.

If particles in the scalar sector are ever discovered, it will be essential to determine their properties, which will give direct information about the nature of the particle and the underlying theory. Three simple examples can be cited:

1. Suppose a Higgs-like particle is discovered with mass 110 GeV. This could either be the standard model (SM) Higgs or an MSSM Higgs. A measurement of the width of the state would presumably tell the difference. However, the SM width is 5 MeV – a formidable measurement!

2. Suppose a Higgs-like particle is discovered with a mass of 150 GeV. This is presumably beyond the MSSM bound, but it could be an NMSSM or an SM Higgs. A measurement of the width could presumably resolve the issue.

3. Suppose a Higgs-like particle of mass 165 GeV is discovered. This is presumably even beyond the NMSSM limits. If this is an SM Higgs, can we learn more by the study of the rare decay modes?

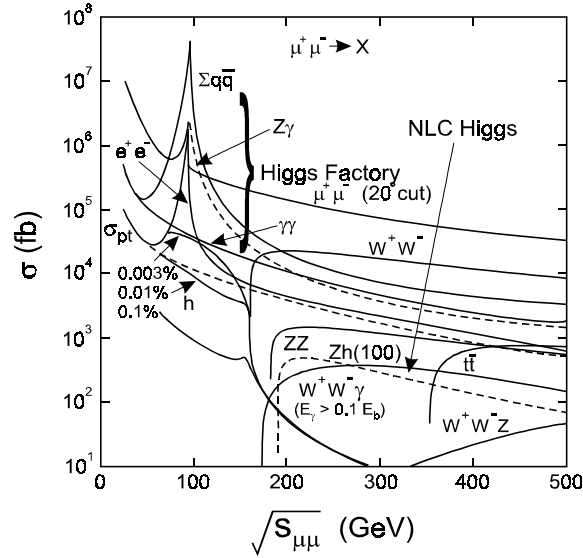

**Figure 2.** The cross sections as a function of energy for $e^+e^-$ and $\mu^+\mu^-$ interactions producing a Higgs boson and other systems.

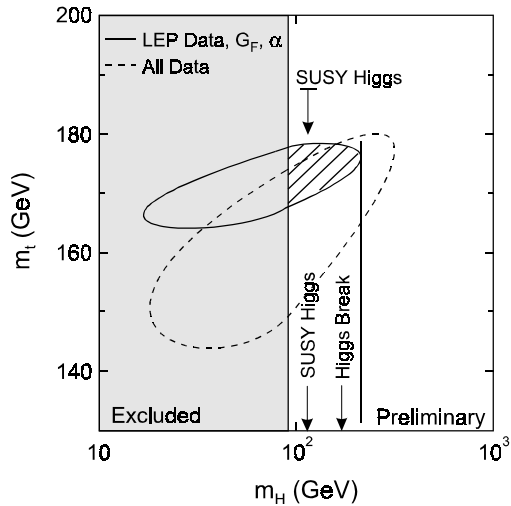

**Figure 3.** The most recent fit of the electroweak parameters from the CERN group to constrain the mass of the Higgs boson.

## THE MYTH OF THE TeV-SCALE SYMMETRY BREAKING

In the old days (e.g., the arguments for the SSC machine), it was argued that "new physics" must appear at the TeV scale. This was based on the concept that a high-mass Higgs (TeV) becomes very "strong" and on the divergent self-interaction of the Higgs sector requiring a cutoff at the TeV scale. However, this need not apply if the Higgs mass is very low, since the stability conditions shown on Fig. 4 allow for a cutoff even up to the Planck mass. The current worldwide fits to all electroweak precision measurements strongly imply that at least one Higgs is low mass. Thus, there may be only one low-mass Higgs to discover at the LHC and nothing more!

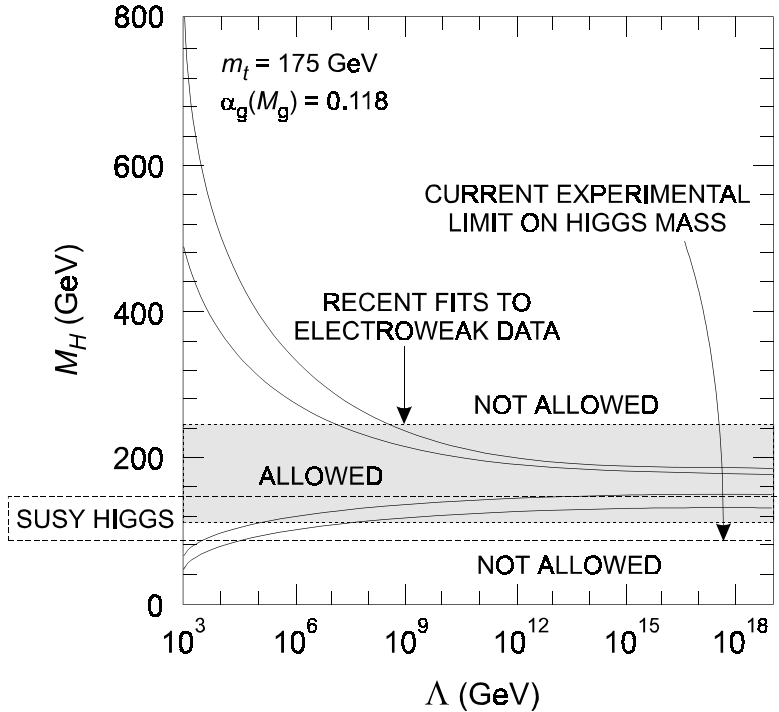

**Figure 4.** Stability diagram for the Higgs scalar at the divergence cutoff Λ and the current values of Higgs mass from electroweak studies.

Take the example of a 170-GeV Higgs. My SUSY theorist friends tell me this is inconsistent with the existence of SUSY (it could also kill string theory). But what cuts off the divergence? The GUT (grand unified theory) scale.

# A NEUTRINO FACTORY AS THE FIRST STAGE OF A HIGGS FACTORY

There is great interest in a neutrino-factory muon storage ring to study neutrino mass. A schematic of such a factory is shown in Fig. 5. We want to point out that this storage ring could be the first stage of a Higgs factory, with the following major differences:
(1) additional cooling is required, especially longitudinal, and
(2) the muon collider storage ring needs to be a small as possible.
There is a certain beautiful symmetry between a study of the origins of mass (Higgs sector) and the study of the tiny neutrino mass.

In the future, we may imagine that the scientific case for the neutrino factory increases and that the technical obstacles are our own – the constraint of a 50-GeV muon storage ring with a relatively small amount of cooling. If the evidence for a low-mass Higgs is confirmed, then the transition from the neutrino factory to a Higgs factory could be fairly straightforward. This could bring a unified approach to the future of particle physics in the USA.

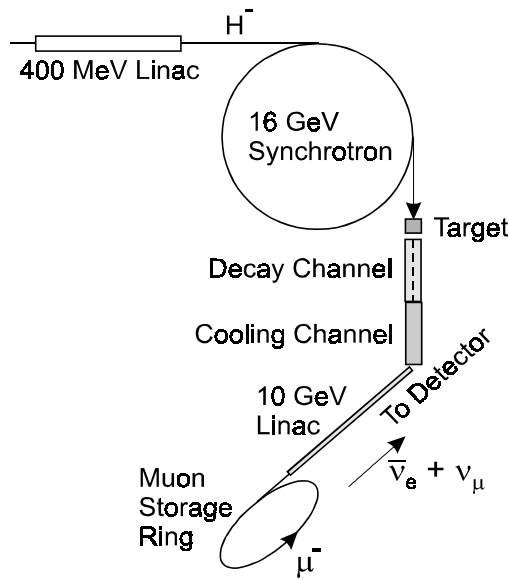

**Figure 5.** Schematic of a neutrino factory at FNAL or BNL.